\documentclass[aps,prc,preprint,superscriptaddress,showpacs]{revtex4}

\usepackage{array,amsmath,graphicx}

\newcommand{\nuc}[3][]{\ensuremath{^{\text{#2}}\text{#3}_{\text{#1}}}}

\begin{document}


\title{New isotope \nuc{44}{Si} and systematics of the production
cross sections of the most neutron-rich nuclei}


\author{O.~B.~Tarasov}
 \email[]{tarasov@nscl.msu.edu}
 \affiliation{National Superconducting Cyclotron Laboratory,
 Michigan State University, East Lansing, MI 48824, USA}
 \affiliation{Flerov Laboratory of Nuclear Reactions,
 JINR, 141980 Dubna, Moscow region,
Russian Federation}
\author{T.~Baumann}
 \affiliation{National Superconducting Cyclotron Laboratory,
 Michigan State University, East Lansing, MI 48824, USA}
\author{A.~M.~Amthor}
 \affiliation{National Superconducting Cyclotron Laboratory,
 Michigan State University, East Lansing, MI 48824, USA}
 \affiliation{Dept.\ of Physics and Astronomy,
 Michigan State University, East Lansing, MI 48824, USA}
\author{D.~Bazin}
 \affiliation{National Superconducting Cyclotron Laboratory,
 Michigan State University, East Lansing, MI 48824, USA}
\author{C.~M.~Folden III}
 \affiliation{National Superconducting Cyclotron Laboratory,
 Michigan State University, East Lansing, MI 48824, USA}
\author{A.~Gade}
 \affiliation{National Superconducting Cyclotron Laboratory,
 Michigan State University, East Lansing, MI 48824, USA}
 \affiliation{Dept.\ of Physics and Astronomy,
 Michigan State University, East Lansing, MI 48824, USA}
\author{T.~N.~Ginter}
 \affiliation{National Superconducting Cyclotron Laboratory,
 Michigan State University, East Lansing, MI 48824, USA}
\author{M.~Hausmann}
 \affiliation{National Superconducting Cyclotron Laboratory,
 Michigan State University, East Lansing, MI 48824, USA}
\author{M.~Mato\v{s}}
 \affiliation{National Superconducting Cyclotron Laboratory,
 Michigan State University, East Lansing, MI 48824, USA}
\author{D.~J.~Morrissey}
 \affiliation{National Superconducting Cyclotron Laboratory,
 Michigan State University, East Lansing, MI 48824, USA}
 \affiliation{Dept.\ of Chemistry, Michigan State University,
 East Lansing, MI 48824, USA}
\author{A.~Nettleton}
 \affiliation{National Superconducting Cyclotron Laboratory,
 Michigan State University, East Lansing, MI 48824, USA}
 \affiliation{Dept.\ of Physics and Astronomy, Michigan State University,
 East Lansing, MI 48824, USA}
\author{M.~Portillo}
 \affiliation{National Superconducting Cyclotron Laboratory,
 Michigan State University, East Lansing, MI 48824, USA}
\author{A.~Schiller}
 \affiliation{National Superconducting Cyclotron Laboratory,
 Michigan State University, East Lansing, MI 48824, USA}
\author{B.~M.~Sherrill}
 \affiliation{National Superconducting Cyclotron Laboratory,
 Michigan State University, East Lansing, MI 48824, USA}
 \affiliation{Dept.\ of Physics and Astronomy, Michigan State University,
 East Lansing, MI 48824, USA}
\author{A.~Stolz}
 \affiliation{National Superconducting Cyclotron Laboratory,
 Michigan State University, East Lansing, MI 48824, USA}
\author{M.~Thoennessen}
 \affiliation{National Superconducting Cyclotron Laboratory,
 Michigan State University, East Lansing, MI 48824, USA}
  \affiliation{Dept.\ of Physics and Astronomy, Michigan State University,
  East Lansing, MI 48824, USA}

\date{\today}

\begin{abstract}
The results of measurements of the production of neutron-rich nuclei by
the fragmentation of a \nuc{48}{Ca} beam at 142~MeV/u are presented.
Evidence was found for the production of a new isotope that is the most
neutron-rich silicon nuclide, \nuc{44}{Si}, in a net neutron pick-up
process.  A simple systematic framework was found to describe the
production cross sections based on thermal evaporation from excited
prefragments that allows extrapolation to other weak reaction products.
\end{abstract}


\maketitle


\section{Introduction\label{Intro}}
The study of properties of the most exotic isotopes continues to be one
of the important tasks in experimental nuclear physics. In addition,
masses, lifetimes, and properties of excited states are important not
only for models of nuclear structure but also for the understanding of
astrophysical processes. The first step in the study of a new exotic
nucleus is its observation, which for neutron-rich nuclei demonstrates
its stability with respect to particle emission.

\begin{figure}
\includegraphics[width=87mm]{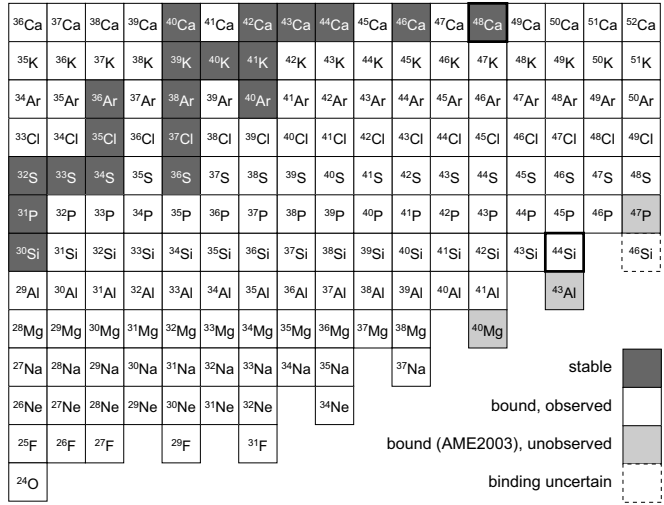}%
\caption{The region of the chart of nuclides under investigation in
this work.\label{chart}}
\end{figure}

The neutron dripline is only confirmed up to $Z=8$ (\nuc[16]{24}{O}) by
work at projectile fragmentation facilities in the US \cite{MF-PRC96},
France \cite{DGM-PRC90,OT-PL97}, and Japan \cite{HS-PLB99}. As
indicated in Fig.~\ref{chart}, the dripline rapidly shifts to higher
neutron numbers at $Z=9$ and \nuc[22]{31}{F} has been observed by
several groups \cite{MN-PLB02,AML-JPG02,EK-AIP07}. This shift makes the
search for the neutron dripline in this region especially difficult but
none the less important.  Experiments at RIKEN in Japan \cite{MN-PLB02}
and at GANIL in France \cite{AML-JPG02} observed the two heaviest
isotopes along the $A=3Z+4$ line, \nuc[24]{34}{Ne} and
\nuc[26]{37}{Na}, by the fragmentation of \nuc[28]{48}{Ca} projectiles.
The heavier nuclei in this series, \nuc[28]{40}{Mg} and
\nuc[30]{43}{Al}, are unobserved at present. All nuclei with $A=3Z+3$
up to $Z=12$ have been shown to be unbound. The neighboring nuclei with
$A=3Z+2$ have been observed up to \nuc[28]{41}{Al} but the production
of the heavier nuclei from a \nuc{48}{Ca} beam requires a reaction with
a net neutron pick-up.

In the present work a series of measurements was carried out to search
for new neutron-rich isotopes in this region and to measure the cross
sections for production of these isotopes. A particular candidate for
study is \nuc[30]{44}{Si}, a nuclide that has two more neutrons than
the projectile nucleus.  Nucleon pick-up products have been observed
among fragmentation products, see for example
Refs.~\cite{GS-PRC92,RP-PRC95,MM-PRC06}, but their cross sections are
significantly lower than those of pure fragmentation processes. The new
data for the production cross sections builds upon the recent results
from Mocko et al.\ \cite{MM-PRC06} and can provide a path to the
production of the most neutron-rich nuclei.

\section{Experimental Details\label{Expt}}
A 142~MeV/u \nuc{48}{Ca} beam from the coupled cyclotron facility at
the National Superconducting Cyclotron Laboratory was used to irradiate
either a \nuc{9}{Be} target (724~mg/cm$^{2}$) or a \nuc{nat}{W} target
(1111~mg/cm$^{2}$) located at the normal target position of the A1900
fragment separator \cite{DJM-NIMA03}. The tungsten target was used due
to its high melting point even though it is not monoisotopic. The
average primary beam intensity for the measurements of the most exotic
fragments was 70~pnA\@. The momentum acceptance of the separator was
either $\Delta p/p = \pm1\%$ or $\pm2\%$ and the angular acceptance was
8~msr. The experimental setup and analysis procedures used for this
experiment were similar to those described in
Refs.~\cite{TB-PRC03,AS-PLB05,MM-PRC06,EK-AIP07} and only the
differences will be briefly described. The time of flight of each
particle that reached the focal plane was measured in two ways: first,
over the 17.8~m flight path between a plastic scintillator
(22~mg/cm$^2$ thick) located at the second dispersive image (image 2)
and a 10~cm thick plastic backstop scintillator located at the focal
plane of the separator, and also over the entire 35.6~m flight path of
the A1900 fragment separator by measuring the arrival time relative to
the phase of the cyclotron rf-signal. The magnetic rigidity for each
particle was determined by the separator setting plus a correction
based on the position measurements at image 2 with the plastic
scintillator, and at the focal plane of the separator using a set of
parallel-plate avalanche counters (PPACs). The standard focal plane
detector setup was augmented to have three silicon PIN diodes
($50\times50$~mm$^2$ by 496~$\mu$m, 528~$\mu$m, and 526~$\mu$m thick)
to enable multiple measurements of the energy-loss of the fragments and
thus provide redundant determinations of the nuclear charge of each
fragment. The simultaneous measurements of multiple $\Delta E$ signals,
the magnetic rigidity, a scintillator signal proportional to the total
energy, as well as the flight times for each particle provided an
unambiguous identification of the atomic number, charge state, and mass
of the produced fragments. The position and angle measurements with
PPACs at the focal plane also enabled discrimination against various
scattered particles.

The relative beam current was monitored continuously by a small BaF$_2$
crystal mounted on a photomultiplier tube near the target position that
provided a normalization for the data obtained at different magnetic
rigidities.  In order to map out the momentum distributions of the
fragmentation products and provide the production yields, the magnetic
rigidity of the separator was varied stepwise from 4.13~Tm to 4.89~Tm.
The momentum distributions of isotopes between magnesium and phosphorus
that were present at these settings were analyzed during the
experiment. These measured distributions are in good agreement with
\textsc{lise++} \cite{OT-NPA04b} calculations, using either the
Universial Parameterization \cite{OT-NPA04} or the model by Morrissey
\cite{DJM-PRC89}, so that the optimum separator setting for the
heaviest isotopes (that were produced at a higher rigidity setting with
very low rates) could be inferred from our \textsc{lise++}
calculations. Once the optimum setting was determined, an achromatic
energy-loss degrader (\nuc{27}{Al}, 151~mg/cm$^{2}$) was inserted at
image 2 of the A1900 separator in addition to the plastic scintillator
to cut down the range of atomic numbers of the fragments reaching the
focal plane.

\begin{figure}
\includegraphics[width=87mm]{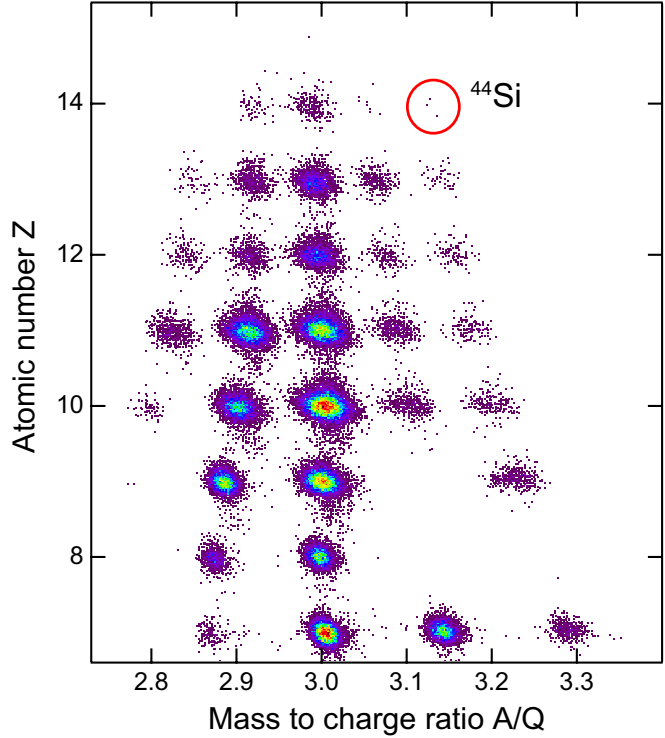}%
\caption{(Color online) Particle identification plot of atomic number $Z$ versus
mass-to-charge ratio $A/Q$ for $Z=7$ to 15.}\label{pid}
\end{figure}

A search for \nuc{44}{Si} was carried out by performing several runs
totaling 4.3 hours with the tungsten target and 5.8 hours with the
beryllium target at a setting optimized for \nuc{38}{Mg} and
\nuc{41}{Al} at a rigidity of 5.045~Tm (4.9598~Tm after image 2).  The
combination of the higher energy loss of silicon isotopes in the thick
targets and the image 2 degrader plus the expected momentum downshift
due to nucleon pickup (cf.\ \cite{RP-PRC95}) placed \nuc{44}{Si} in the
acceptance of the fragment separator. Three events identified as
\nuc{44}{Si} nuclei were observed during the measurements with the
tungsten target (see Fig.~\ref{pid}) and none were observed with the
beryllium target.  The overall efficiency of the system was found to be
$73^{+19}_{-15}$\% and $39^{+19}_{-12}$\% when running with the
tungsten and beryllium targets, respectively.  The efficiency was
dominated by the deadtime of the data acquisition system and the
discrimination against pileup events in the focal plane detector.
Trigger rates averaged 200~Hz for the runs with the tungsten target,
and 450~Hz with the berryllium target. The simulated angular
transmission ranged from 77\% for \nuc{38}{Mg} to 84\% for \nuc{44}{Si}
with an estimated uncertainty of 5\% using the technique described by
Mocko et al.\ \cite{MM-PRC06}.

\section{Results and Discussion\label{Res}}
\begin{figure}
\includegraphics[width=87mm]{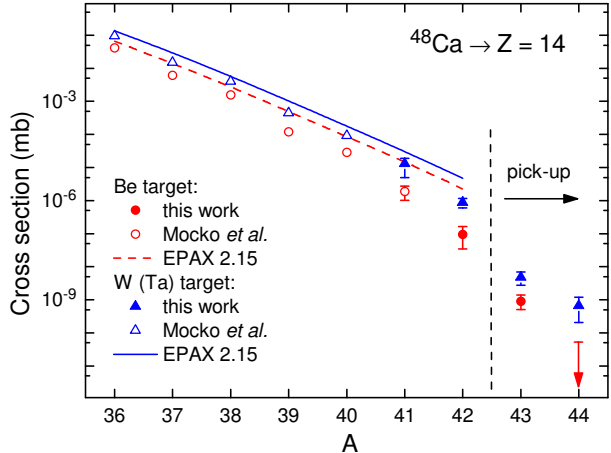}%
\caption{(Color online) The cross sections for production of
neutron-rich silicon nuclei from Ref.~\protect{\cite{MM-PRC06}} and the
present work. The data with $A<43$ (i.e., $N<28$) are compared to the
\textsc{epax} systematics for limiting fragmentation
\protect{\cite{KS-PRC00}}. \label{Z14_cs}}
\end{figure}

The cross sections for the production of neutron-rich silicon isotopes
from this work are shown in Fig.~\ref{Z14_cs} and given in
Table~\ref{Tab_cs} along with the cross sections recently reported by
Mocko et al.\ \cite{MM-PRC06} for the reaction of \nuc{48}{Ca} with
\nuc{9}{Be} and \nuc{181}{Ta} at the same bombarding energy. For the
purpose of comparison we will consider the tantalum and tungsten
targets as equivalent. The cross sections for reaction with the
tungsten target are larger than those with beryllium by factors that
range from approximately 2.5 at $A=38$ to about 9 at $A=42$, values
that become significantly larger than the ratio of the geometric
reaction cross sections $\sigma_{\text{r}}$
\begin{equation*}
 \frac{\sigma_{\text{r}}(\text{W})}{\sigma_{\text{r}}(\text{Be})}
 \sim\frac{\bigl(A^{1/3}(\text{Ca})+A^{1/3}(\text{W})\bigr)^2}
          {\bigl(A^{1/3}(\text{Ca})+A^{1/3}(\text{Be})\bigr)^2} = 2.7\:.
\end{equation*}

The data show a smooth decline with increasing mass number (or neutron
number) up to $A=42$, and then a precipitous drop by about a factor of
110 for the two silicon isotopes with more neutrons than the
projectile. The slope of the data compares well to the \textsc{epax~2.15}
systematics \cite{KS-PRC00} although the data sets lie below the
predictions. The \textsc{epax} parameterization describes the products
of \emph{limiting fragmentation} that occurs at high bombarding
energies and only depends on the sizes of the target and projectile.
Closer comparison of the prediction to the data shows that the cross
sections for \nuc{42}{Si} from both targets are more suppressed than
the average of the lighter isotopes.  This is consistent with the idea
that the most neutron-rich nuclei come from the decay of excited
primary fragments that are themselves even more neutron-rich (and
suppressed by the process of significant neutron transfer from the
target at these bombarding energies due to momentum mismatch).

\begin{table}
 \setlength{\extrarowheight}{2pt}
\caption{Cross sections for neutron-rich Mg and Si isotopes observed in
this work.\label{Tab_cs}}
\begin{ruledtabular}
\begin{tabular}{c@{\extracolsep{\fill}}r@{\extracolsep{0em}}l@{\extracolsep{\fill}}r@{\extracolsep{0em}}l}
 Isotope    & \multicolumn{2}{c}{$\sigma$ (W target)}   &\multicolumn{2}{c}{$\sigma$ (Be target)}   \\
            & \multicolumn{2}{c}{(mb)}                &\multicolumn{2}{c}{(mb)}\\\hline
\nuc{36}{Mg}&$(5\pm1)$      &$\times10^{-6}$            &$(6^{+4}_{-3})$        &$\times10^{-7}$\\
\nuc{37}{Mg}&$(9^{+3}_{-2})$&$\times10^{-8}$           &$(1.6^{+0.8}_{-0.7})$  &$\times10^{-8}$\\
\nuc{38}{Mg}&$(4\pm1)$      &$\times10^{-8}$           &$(4\pm1)$              &$\times10^{-9}$\\[2ex]
\nuc{41}{Si}&$(1.3^{+0.6}_{-0.8})$&$\times10^{-5}$      &&\\
\nuc{42}{Si}&$(9\pm3)$      &$\times10^{-7}$           &$(9^{+7}_{-6})$        &$\times10^{-8}$\\
\nuc{43}{Si}&$(5\pm2)$      &$\times10^{-9}$           &$(9^{+5}_{-4})$        &$\times10^{-10}$\\
\nuc{44}{Si}&$(7\pm5)$      &$\times10^{-10}$           &\\
\end{tabular}
\end{ruledtabular}
\end{table}

\begin{figure*}
\includegraphics[width=\linewidth]{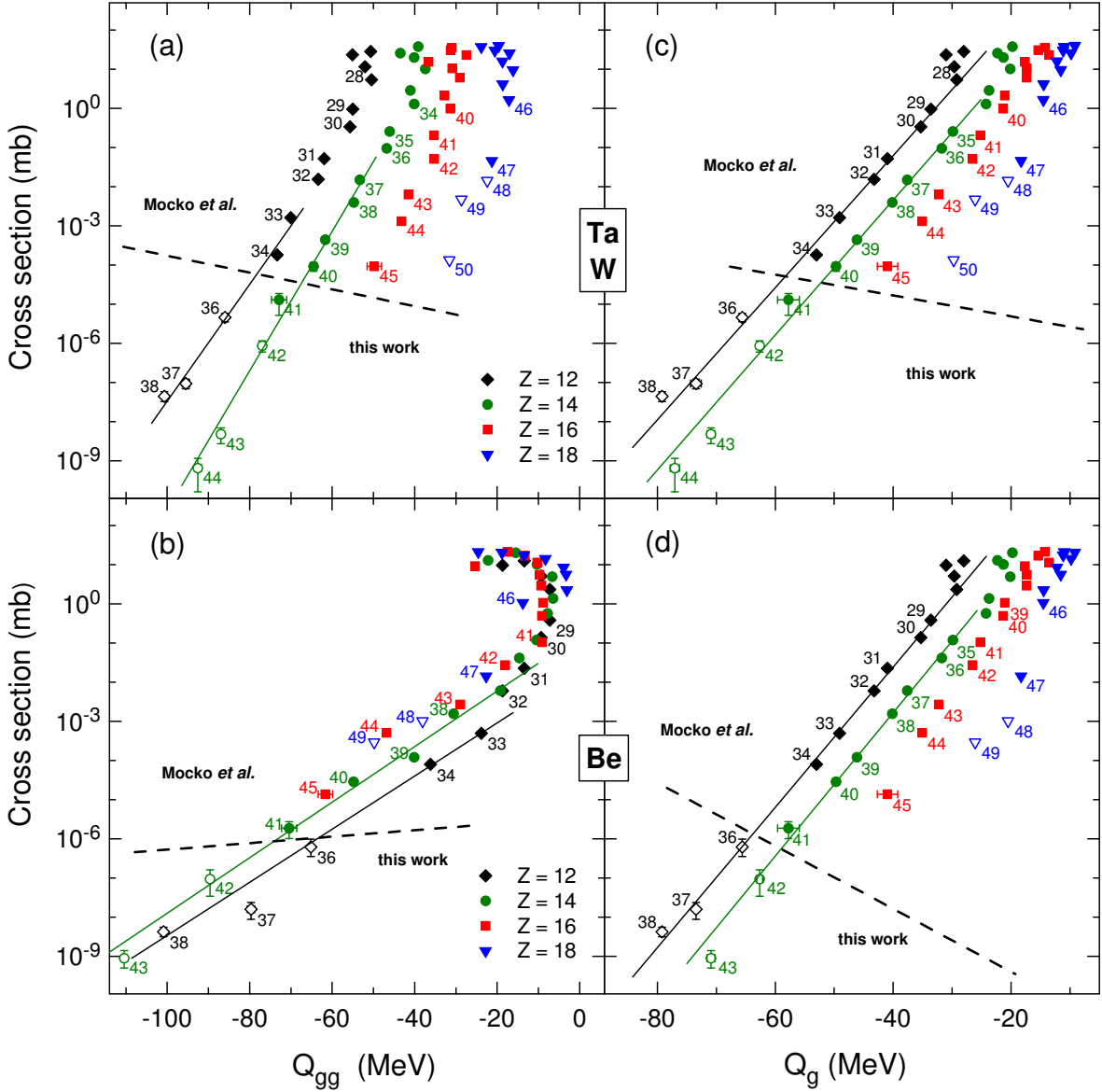}%
\caption{(Color online) The variation of the cross sections for the
production of neutron rich nuclei as a function of the two-body $Q$
values ($Q_{\text{gg}}$, left panels a, b) and as a function of the
one-body $Q$ value ($Q_{\text{g}}$, right panels c, d), see text for
details. Upper panels (a, c) show data for W (Ta), lower panels (b, d)
for Be targets. Each symbol is labeled with the respective mass number.
Data from the present work (below the dashed lines in each panel) were
combined with data from Ref.~\protect{\cite{MM-PRC06}}. Solid symbols
represent $Q$-value calculations based on the measured mass values, and
open symbols based on the recommended values \protect\cite{GA-NPA03,
OT-MSUCL02}. The lines represent exponential fits of the most
neutron-rich isotopes for each chain.\label{Qg_gg}}
\end{figure*}

Models of nuclear reactions used for counting rate estimates, like
the intranuclear-cascade plus evaporation model \cite{DJM-PRL79} or
abrasion-ablation in \textsc{lise++} \cite{OT-NIM2003}
can not reproduce the low yields of the
exotic nuclei observed in this study. Thus it is not possible to
make reliable predictions for further work. As a starting point, the
cross sections in peripheral two-body reactions have been analyzed in
the framework of the $Q_{\text{gg}}$ systematics for a long time
\cite{CKG-PR78}. The central idea of the $Q_{\text{gg}}$ systematics is
that the products are created in a statistical, thermal process and the
cross section should follow the expression
\begin{equation*}
 \sigma(Z,A) = f(Z)\exp{(Q_{\text{gg}}/T)} \quad \text{or} \quad \ln{\bigl(\sigma(Z,A)\bigr)} \propto Q_{\text{gg}}\:,
\end{equation*}
where $Q_{\text{gg}}$ is the simple difference between the mass
excesses  of the ground states of the product and reactant nuclei and
$T$ is an effective temperature that is fitted to the data.  Such an
ansatz is reasonable at low energies when the nuclei undergo a slow
transfer process and for the observation of projectile residues from
mass-asymmetric reactions where the bulk of the excitation energy is
carried by the heavy partner.  Over the years a number of measurements
of light products at higher energies have found some agreement with
this model as can be seen in Fig.~\ref{Qg_gg} (left panels) for the
data from this study combined with the data from Mocko et al.\
\cite{MM-PRC06}. The data for the most neutron-rich isotopes in each
chain tend toward straight lines but the bulk of the data with the
highest yields, highest precision, and lowest $Q$-values behaves very
differently. It is important to realize that $Q_{\text{gg}}$ is most
sensitive to the mass of the lighter fragment since the binding energy
changes most rapidly with neutron and proton number in the low mass
region. Previous studies that were used to develop the $Q_{\text{gg}}$
systematics relied on the analysis of the distributions of the light
fragment from reactions in normal kinematics \cite{CKG-PR78}.  In the
present work the lighter fragment is the \emph{target residue} in the
case of the beryllium target, whereas it is the \emph{projectile
residue} in the case of the tungsten target. The dominant factor in the
exponential is then either the unobserved fragment (beryllium target,
panel (b) of Fig.~\ref{Qg_gg}) or the observed fragment (heavy target,
panel (a) of Fig.~\ref{Qg_gg}).

Projectile fragmentation is usually not described as a
two-body process, but rather as a sudden process that forms an excited
prefragment followed by statistical decay. Charity \cite{RJC-PRC98} has
pointed out that the sequential evaporation of light particles from
sufficiently excited nuclei follows a general pattern that leads to a
somewhat uniform distribution of final products. This uniform
distribution underlies the \textsc{epax} systematics. In the usual case
neutrons are emitted preferentially from excited nuclei until the point
at which the ratio of the widths for statistical emission of neutrons
to charged particles, $\Gamma_N/\Gamma_Z$, becomes small. Note that
this expression includes neutrons and protons bound in clusters as
described in Ref.~\cite{RJC-PRC98}.  The individual emission widths,
$\Gamma_N$ and $\Gamma_Z$, contain a number of factors but most of
these factors approximately cancel in the ratio and the largest
remaining term is an exponential of the difference between the neutron
and proton separation energies, $S_{\text{n}}$ and~$S_{\text{p}}$:
\begin{equation}
\Gamma_N/\Gamma_Z \propto \exp{(S_{\text{p}}-S_{\text{n}})}\:.
\end{equation}
The separation energies contain the masses of the daughter isotopes,
thus, we can expect an exponential dependence of the yield on the mass
difference between the daughter nuclei for proton and neutron emission
in this model. The masses are not known experimentally for most of the
very neutron-rich nuclei in this study. In an attempt to extract the
average systematic behavior the cross sections are plotted as a
function of
\begin{equation}
Q_{\text{g}} = ME(Z=20,A=48)-ME(Z,A)
\end{equation}
in Fig.~\ref{Qg_gg} (right panels), where $ME(Z,A)$ is the mass excess
in MeV. $Q_{\text{g}}$ is a function that compares the relative binding
energies of all of the projectile fragments without regard to the
target nucleus and is a plausible basis for comparison of products from
a process that creates a small set of highly excited nuclei that then
statistically populate all of the available mass surface. The figure
shows that this function provides an excellent systematization of the
data with each isotopic chain falling on a straight line. Moreover, the
slopes or inverse temperatures decrease with atomic number and go from
about 1.2 (Ar from Be) to a maximum of $T\approx2.5$~MeV (Mg and Si
from Be and Ta). The line from the production of magnesium isotopes can
be extrapolated to predict a cross section of $0.04\pm0.01$~pb for
\nuc{40}{Mg}, as yet unobserved.

\section{Summary\label{Sum}}
The study of the production of the most neutron-rich silicon isotopes
provided evidence for the existence of a new isotope, \nuc{44}{Si}, in
a high energy reaction that requires the net transfer of two neutrons
to the projectile.  The decline of the cross sections for the
production of silicon isotopes with increasing mass number was found to
parallel the \textsc{epax} parameterization but at a lower level, up to
the point that neutron pickup intermediates begin to be important. The
measured cross sections for nuclei with more neutrons than the
projectile fall by approximately two orders of magnitude below a
logarithmic extrapolation from the lighter isotopes.

The variation of the cross sections for a large range of reaction
products were considered in the framework of the well-known
$Q_{\text{gg}}$ systematics developed for low-energy two-body
reactions. Only the tails of the distributions had the expected linear
dependence and the applicability of this model to projectile residues
from reverse kinematical reactions is entirely questionable.  On the
other hand, all of the available data were shown to follow a very
smooth systematic dependence, independent of the target, with the mass
of the observed fragment.  An extrapolation of the data using the new
one-body $Q_{\text{g}}$ systematics indicates that a search for
\nuc{40}{Mg} is feasible.

\begin{acknowledgments}
The authors would like to acknowledge the work of the operations staff
of the NSCL to develop the intense $^{48}$Ca beam necessary for this
study.  This work was supported by the U.S.~National Science Foundation
under grant PHY-06-06007.
One of the authors (M.M.) acknowledges support by the U.S.~National Science Foundation
under grant PHY-02-16783 (Joint Institute of Nuclear Astrophysics).
\end{acknowledgments}


\end{document}